\documentclass[12pt,preprint]{aastex}
\shorttitle{}
\shortauthors{ }
\usepackage{graphicx}
\usepackage{subfigure}
\include{eps}

\begin{document}

\title{A comparative study of the spatial distribution of ultraviolet and
  far-infrared fluxes from M~101} 

\author{Cristina C. Popescu\altaffilmark{1}, 
       Richard J. Tuffs\altaffilmark{1},
       Barry F. Madore\altaffilmark{2,3}, A. Gil de Paz\altaffilmark{2},
       Heinrich J. V\"olk\altaffilmark{1},
Tom Barlow\altaffilmark{4},
Luciana Bianchi\altaffilmark{5}, 
Yong-Ik Byun\altaffilmark{6}, 
Jose Donas\altaffilmark{7},
Karl Forster\altaffilmark{4},  
Peter G. Friedman\altaffilmark{4},
Timothy M. Heckman\altaffilmark{8}, 
Patrick N. Jelinsky\altaffilmark{9}, 
Young-Wook Lee\altaffilmark{6}, 
Roger F. Malina\altaffilmark{7}, 
Christopher D. Martin\altaffilmark{4}, 
Bruno Milliard\altaffilmark{7}, 
Patrick Morrissey\altaffilmark{4},
Susan G. Neff\altaffilmark{10}, 
R. Michael Rich\altaffilmark{11}, 
David Schiminovich\altaffilmark{4}, 
Oswald H. W. Siegmund\altaffilmark{9},
Todd Small\altaffilmark{4},
Alex S. Szalay\altaffilmark{8},
Barry Y. Welsh\altaffilmark{9},
Ted K. Wyder\altaffilmark{4}
}  
\altaffiltext{1}{Max Planck Institut f\"ur Kernphysik, Saupfercheckweg 1, 
69117 Heidelberg, Germany} 
\altaffiltext{2}{The Observatories of the Carnegie Institution of Washington, 
813 Santa Barbara Str., Pasadena, 91101 California, USA}
\altaffiltext{3}{NASA/IPAC Extragalactic Database, 770 S. Wilson Avenue,
  Pasadena, California 91125, USA}
\altaffiltext{4}{California Institute of Technology, MC 405-47, 1200 East
California Boulevard, Pasadena, CA 91125}

\altaffiltext{5}{Center for Astrophysical Sciences, The Johns Hopkins
University, 3400 N. Charles St., Baltimore, MD 21218}

\altaffiltext{6}{Center for Space Astrophysics, Yonsei University, Seoul
120-749, Korea}

\altaffiltext{7}{Laboratoire d'Astrophysique de Marseille, BP 8, Traverse
du Siphon, 13376 Marseille Cedex 12, France}

\altaffiltext{8}{Department of Physics and Astronomy, The Johns Hopkins
University, Homewood Campus, Baltimore, MD 21218}

\altaffiltext{9}{Space Sciences Laboratory, University of California at
Berkeley, 601 Campbell Hall, Berkeley, CA 94720}

\altaffiltext{10}{Laboratory for Astronomy and Solar Physics, NASA Goddard
Space Flight Center, Greenbelt, MD 20771}

\altaffiltext{11}{Department of Physics and Astronomy, University of
California, Los Angeles, CA 90095}

\begin{abstract}
The total ultraviolet (UV) flux (from 1412 to 2718\,\AA) of M~101 is compared 
on a pixel-to-pixel basis with the total far-infrared (FIR) flux (from 60 to
170\,${\mu}$m), using the maps of the galaxy taken by GALEX in the near-UV and
far-UV and by ISOPHOT at 60, 100 and 170\,${\mu}$m. The main result of 
this investigation is the discovery of a tight dependence of the FIR/UV 
ratio on radius, with values monotonically decreasing from $\sim 4$ in the 
nuclear region to nearly zero towards the edge of the optical disk. 
  Although the tightness of this dependence is in part attributable to
  resolution effects, the result is 
consistent with the presence of a large-scale distribution of diffuse dust 
having a face-on optical depth which decreases with radius and which dominates
 over the more localized variations in opacity between the arm and 
interarm regions. We also find a trend for 
the FIR/UV ratio to take on higher values in the regions of diffuse interarm
emission than in the spiral-arm regions, at a given radius. This is interpreted
quantitatively in terms of the escape probability of UV photons from spiral
arms and their subsequent scattering in the interarm regions, and in terms of 
the larger relative contribution of optical photons to the heating of the dust
 in the interarm regions.
\end{abstract}

\keywords{galaxies: individual (M 101)---galaxies: spiral---ultraviolet: galaxies---infrared: galaxies---(ISM:) dust, extinction---scattering}

\section{Introduction}

It is not known to which extent the appearance of ultraviolet (UV) 
images of gas-rich star-forming galaxies differs from the intrinsic 
distribution of UV sources, due to the effects of absorption and 
scattering by dust grains. These effects can be quantified by a direct 
comparison of UV maps with maps of Far-Infrared (FIR) emission from the 
grains, since most of the absorbed UV light is re-radiated in the FIR. 
Late-type face-on spiral galaxies are ideal for such studies because a higher 
proportion of their bolometric output originates from the young stellar
population emitting in the UV, and because the analysis is not complicated
by inclination effects.

The nearly face-on Sc galaxy M~101 was observed by GALEX 
(Galaxy Evolution Explorer; Martin et al. 2004, this volume) as part of the
GALEX Nearby Galaxies Survey (NGS). The GALEX UV
photometry of all discrete sources in M~101 is presented by Bianchi et
al. (2004, this volume). 
Here we compare the GALEX images of M~101 with maps of 
this galaxy made using the ISOPHOT instrument (Lemke et al. 1996) on board 
the Infrared Space Observatory (ISO). An alternative 
study of another galaxy from the NGS, M83, is presented by 
Boissier et al. (2004, this volume), who derive extinction radial profiles of 
that galaxy from GALEX UV imaging. Extinction radial profiles of a few spiral 
galaxies were also presented by Boissier et al. (2004) using FOCA and IRAS 
data. For statistical samples, Buat et al. (2004, this volume) has presented 
a complementary
study of extinction based on GALEX and IRAS data. Studies of the stellar
populations in the inner and outer disks of NGS galaxies are also presented by
Thilker et al. (2004a,b, this volume).

\section{Comparison between the GALEX and ISOPHOT images}

GALEX observed M~101 in its far-UV (FUV; 1530\,\AA) and near-UV
(NUV; 2310\,\AA) bands (Morrissey et al. 2004, this volume).
Using the GALEX pipeline (Martin et al. 2001), which 
final images were produced with a spatial scale of 
1.5\,$^{\prime\prime}$\,pixel$^{-1}$. The PSF
FWHM of the images were $\sim$4 and 5\,$^{\prime\prime}$ for the FUV and NUV
bands, respectively. 
The ISOPHOT images were made in bands centered at 60, 100 and 170\,${\mu}$m,
covering an overall spectral range from 40 to 240\,${\mu}$m. 
The FWHM of Gaussian beams having the same area as the ISOPHOT beams are 
50.5, 54.5 and $107.3^{\prime\prime}$ at 60, 100 and 170\,${\mu}$m, 
respectively. Details 
of the data analysis for the ISOPHOT observations of M~101 are given in 
Tuffs \& Gabriel (2003). 

To compare the GALEX maps with the ISOPHOT maps we converted the 
UV maps to the orientation, resolution and sampling of the FIR maps. 
The UV images were convolved with the ISO beams and resampled every 
$15.33^{\prime\prime}\times 23.00^{\prime\prime}$ for
comparison with the 60 and 100\,${\mu}$m images, and every
$30.66^{\prime\prime}\times 92.00^{\prime\prime}$ for comparison with the
170\,${\mu}$m image. The resulting images were corrected for
Galactic extinction and then combined by
means of a linear interpolation and integrated over wavelength to produce
images of the UV flux integrated from 1412\,\AA\ to 
2718\,\AA, calibrated in W/m$^2$. We refer to these combined images as 
``total UV'' images, even though they do not contain the emission between
912\,\AA\ and 1412\,\AA\ and between 2718\,\AA\ and the U band. For a
steady-state star-formation and a Salpeter IMF, we estimate that the
factor by which one must multiply the emission in the range $1412 -
2718$\,\AA\ to obtain the true total UV emission is 1.95. This factor
depends only to second order on the effects of reddening. 

To facilitate a quantitative comparison between the ``total UV'' and FIR 
images, the ISOPHOT images were converted into units of 
W/m$^2$ by multiplying them by the corresponding filter widths.
In the top panels of Fig.~1 we display as an example the 
filter-integrated 100\,${\mu}$m ISOPHOT image (left) together with the
corresponding ``total UV'' image (right). The 100\,${\mu}$m 
image appears smaller than the UV image, mainly because the FIR counterparts to
the upper spiral arms 
prominent in the UV image are very faint at 100\,${\mu}$m. This 
effect is further quantified by the ratio image (100\,${\mu}$m/UV) displayed 
in the bottom left panel of Fig.~1, where 
the region of the upper spiral arms coincides with low values of this ratio. 
At high surface brightness levels however, the 100\,${\mu}$m image and the
UV image appear to trace similar structures. The prominent HII regions are 
seen in both direct UV
light and in the dust re-emission, albeit with varying ratios. The same is true
for the general spiral structure. Also, in both the 100\,${\mu}$m image and 
the UV image a 
diffuse emission underlies the spiral structure. In order to compare the 
100\,${\mu}$m/UV color from the 
spiral structure with that from the underlying diffuse emission we produced 
an image of the ``spiral arm fraction'', whose values give the fraction of the 
beam area occupied by spiral arm structure. This was generated from the high 
resolution UV image 
and is displayed in the bottom right panel of Fig.~1. Comparison between the 
ratio
image and the ``spiral arm fraction'' image shows that the high values of the 
100\,${\mu}$m/UV ratio trace the interarm regions. In other words the 
``spiral features'' in the ratio image are in reality regions of diffuse 
emission which are interspaced with the real spiral features, as seen in the 
``spiral arm fraction'' image.

\section{The radial dependence of the FIR/UV ratio}

A fundamental property of galaxies is the fraction of light from
young stars which is re-radiated by dust. With the advent of GALEX and FIR
facilities like ISO, one can not only investigate this property for the 
spatially integrated emission (Xu \& Buat 1995, Popescu \& Tuffs 2002)
but also as a function of position in the galaxy.  To obtain a ``total FIR'' 
image we combined the ISOPHOT 60, 100 and 170\,${\mu}$m images (at the
resolution of the 170\,${\mu}$m image) by linearly 
interpolating between the bands and integrating over wavelength (between 60 and
170\,${\mu}$m), analogous to the procedure adopted to
obtain the ``total UV'' image. 
As shown by Bothun \& Rogers (1992), the dust in the inner regions of
  M~101 is warmer than the dust in the outer regions and there is 
considerable variation in the 60/100
${\mu}$m ratio throughout the face of the galaxy indicating a wide range of 
heating conditions. However, the FIR luminosity derived here is little 
influenced by dust
temperature (or emissivity) variations, since
the peak
of all emission components from the warm and cold dust should lie within
the broad spectral range of our filters (40-240\,${\mu}$m).
The 170\,${\mu}$m band is particularly important in measuring the cold dust
which accounts for most of the dust mass as well as bulk of the dust 
luminosity (Tuffs \& Popescu 2003). 
The derived ``total FIR'' image still does not contain 
grain emission in the submm and MIR spectral ranges. 
Correcting for this spectral incompleteness in
the same way as done by Popescu et al. (2002) for the late-type Virgo Cluster 
galaxies observed by ISOPHOT (Tuffs et al. 2002a,b) in the same filters as 
M~101, we obtain a correction factor of about 2.  This is 
comparable to the corresponding correction factor
to convert fluxes in the ``total UV'' GALEX band into the true total UV fluxes.
Thus the ratio between the ``total FIR'' and ``total UV'' fluxes should be
comparable to the ratio between the total flux from grains and the UV flux
which would have been observed from 912\,\AA\ to the U band. 

By dividing the ``total FIR'' image with the ``total UV'' image we
obtained a ratio FIR/UV map with a radial dependence as depicted in
Fig.~2. In the derivation of the radial distance we neglected the small 
inclination ($i=18^{\circ}$; Sofue et al. 1997), since the value of the 
position angle is not well constrained and the effect on the result is 
very small. Fig.~2 shows
a remarkably tight dependence of the FIR/UV ratio on radius, with a monotonic 
decrease from values approaching 4 in the nuclear region to
values approaching zero in the outermost regions. The tightness of the
dependence is presumably, at least in part, an effect of the large beam which
averages the emission from sources of different FIR/UV colours. Nevertheless,
some points of high FIR/UV values emerge above the general trend between radii 
of 200$^{\prime\prime}$ and 500$^{\prime\prime}$.
These points originate from an interarm region 
to the SE of the nucleus. If these exceptional points are neglected,
the radial variation of the FIR/UV ratio is well fitted by an offset 
exponential $f(r) = a(0) * exp(-r/a(1)) + a(2)$ with $a(0) = 3.98\pm 0.19$, 
$a(1) = 360.0^{\prime\prime}\pm 43.7^{\prime\prime}$ and 
$a(2) = -0.26\pm 0.19$.

Clearly the major factor determining the FIR/UV ratio is radial position.
To statistically investigate the extent to which the ratio also varies when 
moving from the spiral arms into the interarm region at fixed radius,
we divided the points into radial bins of width 100$^{\prime\prime}$ and, 
within each 
radial bin, we identified the 10$\%$ of points (plotted in red) with the 
highest spiral arm fraction. The latter was calculated by creating an image of
the ``spiral arm fraction'' at the resolution of the 170\,${\mu}$m image,
using an analogous procedure to that described in the previous section.
Over the entire radial range the red points tend to be clustered at the low 
values of the FIR/UV ratio, meaning that this ratio takes
systematically higher values for the diffuse interarm emission than for the 
spiral arm emission at a given radius.

\section{Discussion}

To explain our results we must consider the
sources and the propagation of the UV photons within the galaxy and the
connection of these to the heating and distribution of the dust grains. Here we
consider an approximate analytical treatment of these effects in order 
to indentify the primary reasons for the observed trends. For a more detailed
investigation fully self-consistent radiative transfer techniques would need to
be applied, such as those by Bianchi et al. (1996), Ferrara et al. (1999), 
Baes \& Dejonghe (2001), Tuffs et al. (2004).

In the spiral arm regions the UV photons originate from young stars embedded in
the HII regions. A fraction $F$ of the UV luminosity $L_{\star}$ can be 
considered to be locally absorbed and re-radiated by dust in the vicinity of 
the HII region and a fraction $(1-F)$ escapes the HII 
region. Thus we can treat the HII regions as sources in the spiral arms 
emitting both UV and FIR, with luminosities  
$(1-F)\,L_{\star}$ and $F\,L_{\star}$, respectively. 
The factor F is simply a geometrical blocking factor which provides a ``grey''
attenuation, independent of wavelength. A detailed physical
description of the factor F is given in Tuffs et al. (2004).

Having escaped from the HII regions, the UV light emitted in the
direction of the observer (perpendicular to the disk for a face-on system like
M~101) further passes through a layer of diffuse dust which also attenuates 
the UV and augments the FIR. So the FIR and UV luminosity observed along each 
line of sight will be:
$L_{\rm FIR}^{\rm obs, arm} = F\,L_{\star} + (1-F)\,L_{\star}\,G$
and 
$L_{\rm UV}^{\rm obs, arm} = (1-F)\,L_{\star}\,(1-G)$,
where $G$ is the probability of absorption in the layer of diffuse dust for a
UV photon traveling perpendicular to the disk. The resulting FIR/UV ratio seen
towards the spiral arm regions is then given by: 
\begin{eqnarray}
\frac{\displaystyle L_{\rm FIR}^{\rm obs, arm}}{\displaystyle L_{\rm UV}^{\rm
    obs, arm}}
& = & \frac{\frac{\displaystyle F}{\displaystyle 1-F} + G}{1-G}
\end{eqnarray}
The interarm regions are pervaded by diffuse UV light, with relatively 
few stellar sources (see Bianchi et al. 2004, this volume; Thilker et
al. 2004a,b this volume).
 Because of the smooth appearance of the diffuse interarm emission, we will 
work under the hypothesis that this diffuse interarm UV emission is 
radiation escaping from the spiral arms, traveling in the plane of 
the disk and subsequently being scattered by grains into the observer's line 
of sight
perpendicular to the disk. Thus we can treat the dust grains scattering the
 UV light as ``sources'' in the interarm regions emitting both UV and FIR,
 with luminosities proportional to $u_{\rm UV}\,\sigma_{\rm ext}\,a_{\rm
   UV}^{\bot}$, and $u_{\rm UV}\,\sigma_{\rm ext}\,(1-a_{\rm UV})$,
 respectively. Here $u_{\rm UV}$ is the local energy density of the UV
 radiation field, $\sigma_{\rm ext}$ is the extinction cross-section of the
 grain, $a_{\rm UV}$ is the angle-averaged albedo of the grain, and 
$a_{\rm UV}^{\bot}$ is the albedo multiplied by the phase function for light 
scattered at $90^{\circ}$. The UV scattered light traveling towards the
observer must then pass perpendicularly through the layer of diffuse dust 
which attenuates the UV and augments the FIR. If we also consider that in
general some fraction ($\eta$) of the FIR emission will be powered by
optical photons, the FIR and UV luminosity 
observed along each line of sight towards the interarm regions  will be:
$L_{\rm FIR}^{\rm obs, inter} \sim 
[u_{\rm UV}\,\sigma_{\rm ext}\,(1-a_{\rm UV}) +
u_{\rm UV}\,\sigma_{\rm ext}\,a_{\rm UV}^{\bot}\,G]/[1-\eta]$ and
$L_{\rm UV}^{\rm obs, inter} \sim u_{\rm UV}\,\sigma_{\rm ext}\,a_{\rm UV}^{\bot}\,(1-G)$.
The resulting FIR/UV ratio seen towards the interarm regions is then given by: 
\begin{eqnarray}
\frac{\displaystyle L_{\rm FIR}^{\rm obs, inter}}{\displaystyle L_{\rm
      UV}^{\rm obs, inter}}
      & = & \frac{\frac{\displaystyle 1-a_{\rm
      UV}}{\displaystyle a_{\rm UV}^{\bot}} + G}{(1-\eta)(1-G)}
\end{eqnarray}

For simplicity we first make the approximation that $\eta=0$, in which case
Eq. (1) has the same functional form with respect to G as Eq. (2). Since 
there is no
reason why $F$ or $a_{UV}$ should depend on radial position, the strong radial
dependence seen in Fig.~2 can only be attributable to the radial
dependence of the factor $G$, the absorption probability for UV photons 
travelling through the diffuse dust. In other words our results imply the 
presence of
a large scale distribution of diffuse dust having a face-on optical depth that
decreases with radius and which dominates local variation in opacity between 
the arm and interarm regions.

For the case $\eta=0$, Eqs. (1) and (2) also indicate that the observed 
systematic
difference between the FIR/UV ratio in the arm and interarm regions is
attributable to the difference between the factors:
$F/(1-F)$ (for the arms) and
$(1-a_{\rm UV})/a_{\rm UV}^{\bot}$ (for the interarm regions), with the amplitude of
the difference also depending on the value of $G$.
Using the values of the albedo given by the model of Laor \& Draine (1993) and
the phase function from Henyey \& Greenstein (1941) we obtain 
$(1-a_{\rm UV})/a_{\rm UV}^{\bot}=2.98$. Values of typically 0.25 for the $F$
factor have been derived from self-consistent modeling of the UV/FIR/submm 
SEDs of normal galaxies (Popescu et al. 2000, Misiriotis et al. 2001), yielding
a value of 0.33 for $F/(1-F)$. These values for $F/(1-F)$ and
$(1-a_{\rm UV})/a_{\rm UV}^{\bot}$ are consistent with the
observed FIR/UV ratio being smaller in the arm than in the interarm region. 

For the case $\eta>0$ the FIR/UV ratio in the interarm region will be further
boosted, due to the expected increase in the fraction of FIR emission
powered by optical photons at larger distances from the HII regions in
the spiral arms. The combined effect of the optical heating and the scattering
of the UV emission means that the FIR/UV ratio will not be a good indicator of
extinction in the interarm region. One should also bear in mind that the 
observed difference in the 
FIR/UV ratio between arm and interarm region from Fig.~2 is in fact reduced 
from 
the prediction of Eqs. (1) and (2) (for any plausible value of $G$) due to 
beam smearing. Furthermore, even in the
most extreme interarm regions it is apparent from the full resolution GALEX
image that some sources of UV emission are present (see Bianchi et al. 2004,
this volume).

\acknowledgments
 
GALEX (Galaxy Evolution Explorer) is a NASA Small Explorer, launched in 
April 2003. We gratefully acknowledge NASA's support for construction, 
operation, and science analysis for the GALEX mission,
developed in cooperation with the Centre National d'Etudes Spatiales
of France and the Korean Ministry of Science and Technology.

\begin{figure}[htb]
\includegraphics[scale=0.8]{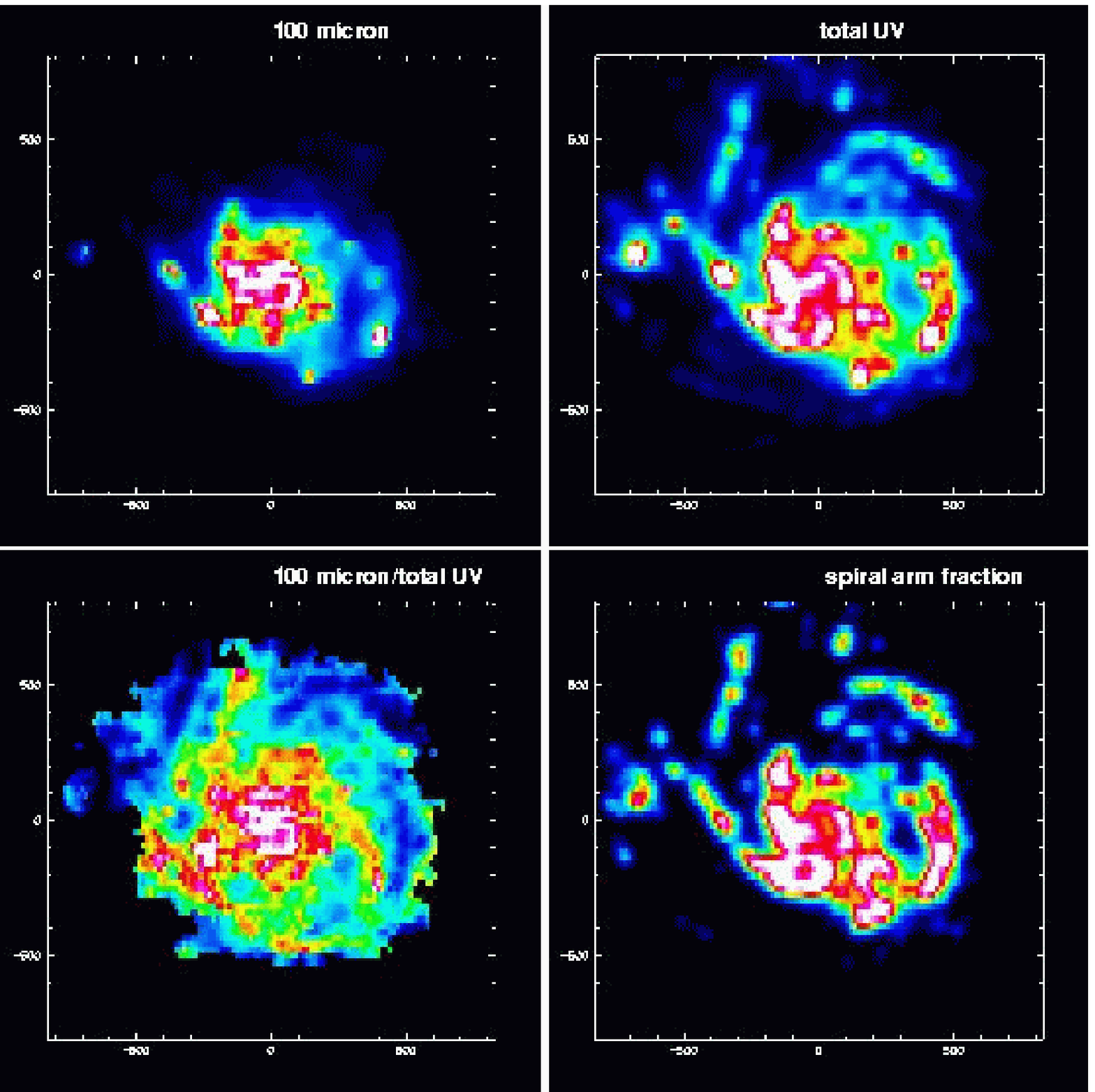}
\caption{Top left panel:filter-integrated 100\,${\mu}$m ISOPHOT image. The 
white color is for fluxes 
$>0.50\times 10^{-14}\,{\rm W}/{\rm m}^2$/pixel and the blue color has typical values of
$\sim 0.06\times 10^{-14}\,{\rm W}/{\rm m}^2$/pixel. The maximum flux is 
$1.38\times 10^{-14}\,{\rm W}/{\rm m}^2$/pixel. Top right panel: ``total UV'' image
converted to the orientation, resolution and sampling of the 100\,${\mu}$m
ISOPHOT image. The white color is for 
fluxes $>0.39\times 10^{-14}\,{\rm W}/{\rm m}^2$/pixel and the blue color has typical 
values of $\sim 0.048\times 10^{-14}\,{\rm W}/{\rm m}^2$/pixel. The maximum flux is 
$1.03\times 10^{-14}\,{\rm W}/{\rm m}^2$/pixel. Bottom left panel: the ratio image of
the filter-integrated 100\,${\mu}$m ISOPHOT image divided by the corresponding
``total UV'' image. The white color is for 
ratios $>1.49$ and the blue color has typical values of $\sim 0.17$. The 
maximum ratio is $2.53$. Bottom right panel: the image of the ``spiral arm
fraction'' at the
orientation, resolution and sampling of the 100\,${\mu}$m ISOPHOT image. The 
white color is for fractions 
$>0.67$ and the blue color has typical values of $\sim 0.08$. The 
maximum fraction is $1.00$. All panels depict a field of 27.7$^{\prime}\times
27.1^{\prime}$ centered at $\alpha^{2000}=14^{\rm h}03^{\rm m}13.11^{\rm s}$;
$\delta^{2000}=54^{\circ}21^{\prime}06.6^{\prime\prime}$. The pixel size is
$15.33^{\prime\prime}\times 23.00^{\prime\prime}$.}
\end{figure}
\begin{figure}[htb]
\includegraphics[scale=0.8]{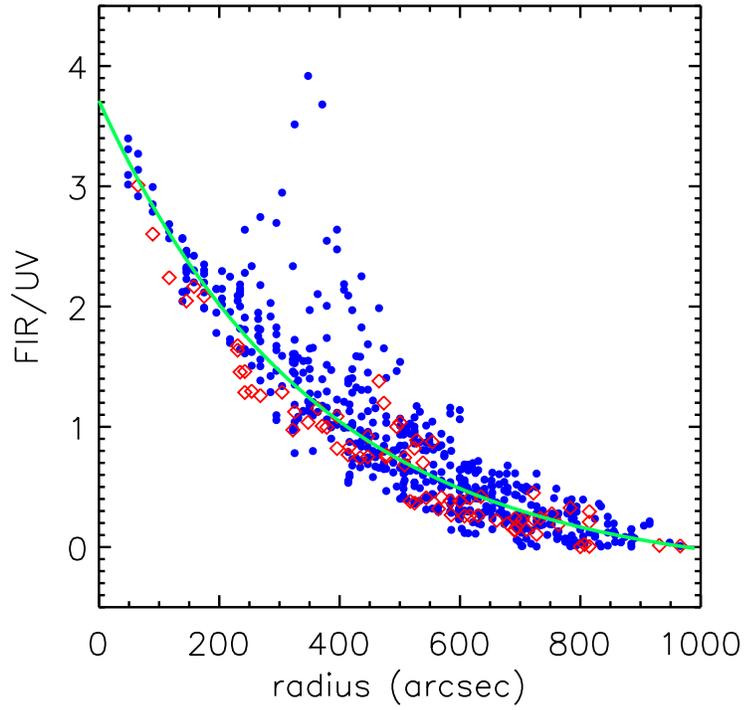}
\caption{The pixel values of the FIR/UV ratio map at the resolution of the 
170\,${\mu}$m image versus angular radius. The blue dots are for lines of
sight towards interarm regions and the red diamonds towards the spiral arm
regions. The green solid line is an offset exponential fit to the data.}
\end{figure}

\end{document}